\documentclass[aps,prd,twocolumn,groupedaddress]{revtex4-1}
\usepackage[dvipdfmx]{graphicx}  % needed for figures
\usepackage{dcolumn}   % needed for some tables
\usepackage{bm}        % for math
\usepackage{amssymb}   % for math
\usepackage{amsmath}	% for Environment split
\usepackage[dvipdfmx]{color}	%図のため
\usepackage{ulem}  % \sout
\newcommand{\eref}[1]{(\ref{#1})}

\newcommand{\dev}[0]{\mathrm d}%微積分のd
\newcommand{\bibun}[3]{\frac{\dev^{#3}{#1}}{\dev{#2}^{#3}}}%微分
\newcommand{\henbibun}[3]{\frac{\partial^{#3}{#1}}{\partial {#2}^{#3}}}
\newcommand{\im}[0]{i}

\newcommand{\Kerr}[0]{{a}}

\newcommand{\diffl}[0]{\pmb{D}}
\newcommand{\diffn}[0]{\pmb{\Delta}}
\newcommand{\diffm}[0]{\pmb{\delta}}
\newcommand{\diffmcc}[0]{\overline{\diffm}}

\newcommand{\rhocc}[0]{\overline{\rho}}

\newcommand{\betacc}[0]{\overline{\beta}}

\newcommand{\pai}[0]{\varpi}
\newcommand{\paicc}[0]{\overline{\varpi}}
\newcommand{\taucc}[0]{\overline{\tau}}
\newcommand{\Psicc}[0]{\overline{\Psi}}

\numberwithin{equation}{section}
\renewcommand\theequation{\arabic{section}.\arabic{equation}}

\newcommand{\mass}[0]{m}
\newcommand{\pusai}[1]{\psi}

\renewcommand{\Re}[0]{{\rm Re}}
\renewcommand{\Im}[0]{{\rm Im}}

\begin{document}

%\preprint{}

\title{Gravitational perturbation induced by a rotating ring around a Kerr black hole}


\author{Yasumichi Sano and Hideyuki Tagoshi
}
%\thanks{}
\affiliation{Department of Earth and Space Science, Graduate School of Science, Osaka University, Toyonaka, Osaka 560-0043, Japan}


\date{\today}

\begin{abstract}
The linear perturbation of a Kerr black hole induced by a rotating massive circular ring 
is discussed by using the formalism by Teukolsky, Chrzanowski, Cohen and Kegeles. 
In these formalism, the perturbed Weyl scalars, $\psi_0$ and $\psi_4$, are first obtained 
from the Teukolsky equation. The perturbed metric is obtained in a radiation gauge 
via the Hertz potential. 
The computation can be done in the same way as in our previous paper \cite{SanoTagoshi14}, 
in which we considered the perturbation of {a} Schwarzschild 
black hole {induced} by a rotating ring. 
By adding lower multipole modes such as mass and angular momentum perturbation 
which are not computed by the Teukolsky equation, and by appropriately setting the parameters 
which are related to the gauge freedom, we obtain the perturbed gravitational field which is smooth 
except on the equatorial plane outside the ring. 
\end{abstract}

%\keywords{}

\maketitle


\section{Introduction}

The black hole perturbation method is important to investigate the physical property of black holes, 
such {as} the quasinormal modes, and the various astrophysical phenomenon 
like the orbital evolution of stars around much larger black holes and the 
the gravitational waves induced by them. 
Especially the later phenomenon include the extreme mass ratio inspirals 
which are one of the most important source{s} for the future space laser interferometers, 
eLISA \cite{eLISA}, DECIGO \cite{SNK, DECIGO}, and BBO \cite{BBO}. 

When the black holes are non-rotating, the metric perturbation can be analyzed 
by using the Regge-Wheeler and the Zerilli equations \cite{RW1957, Z1970},
{which} are the single, decoupled equation{s} for 
the odd and even parity modes, respectively. 
On the other hand, if the black holes are rotating, there is no such formalism.
The perturbation of Kerr black holes are usually analyzed by using the 
Teukolsky equation which is the equation {for} the perturbation of the Weyl scalars, $\psi_0$ and $\psi_4$.
There is a formalism, by Chrzanowski \cite{c} and Cohen and Kegeles \cite{kc, ck}, 
(see also \cite{wald1978, Stewart1979}),
to compute the metric perturbation from the perturbation of $\psi_0$ and $\psi_4$ 
obtained with the Teukolsky equation (hereafter, this is called the CCK formalism). 
In this method, a radiation gauge is used to calculate the metric perturbation. 
Recently, this method becomes more important, mainly because of the necessity to compute 
the gravitational self-force on the point particle orbiting around a Kerr black hole, 
and several works have been done 
\cite{LoustoWhiting, ori03, YunesGonzalez, kfw, BarackOri2001, ksf+10, skf+11, pmb}. 

In spite of these development{s}, there are still only a few examples 
of the explicit computation of the metric perturbation by using the CCK formalism. 
In our previous paper \cite{SanoTagoshi14} (hereafter, Paper I), 
we computed the metric perturbation {induced by} a rotating circular mass ring on the equatorial plane 
around a Schwarzschild black hole. 
We found that if we consider only modes which can be derived from $\psi_0$ or $\psi_4$
(which contain harmonic modes of $l\geq 2$), 
there appear unphysical discontinuities in the Hertz potential and the metric perturbation at the ring radius. 
This discontinuities, however, can be removed by adding the lower modes of $l=0$ and $1$, and 
by setting free parameters, corresponding to the gauge freedom, appropriately. 
We obtained metric perturbation which is regular everywhere except for the equatorial plane 
outside the ring radius. 

In this paper, we extend this analysis to the case of {the} Kerr black hole. 
As in the Schwarzschild case, this problem is stationary and axisymmetric. 
Nevertheless, this problem contains both the mass and the angular momentum perturbation. 
We find that, although some equations must be rederived by using the Kerr metric, 
we can obtain the metric perturbation 
in the same {way} as in the Schwarzschild case in Paper I. 

This paper is organized as follows.
{
In Sec. \ref{section:TeuEq}, we derive 
the perturbed Weyl scalars $\psi_0$ and $\psi_4$ by solving the Teukolsky equation.
In Sec. \ref{sec:construction}, we describe how to obtain the perturbed gravitational field. 
In Sec. \ref{sec:hertz}, 
we obtain the particlular solution of the Hertz potential from $\psi_0$ and $\psi_4$. 
The determination of the homogeneous part of the Hertz potential which contains 
the lower modes $l=0,~1$ is also discussed. 
In Sec. \ref{sec:results}, the perturbed Weyl scalars and metric perturbation are shown. 
Sec. \ref{sec:summary} is devoted to summary and discussion.
}

\section{Solutions of the Teukolsky equation}
\label{section:TeuEq}
%First we analytically derive $\psi_0$ and $\psi_4$. 
The Kerr metric is given in Boyer-Lindquist coordinates as 
\begin{equation}
\begin{split}
\dev s^2 
&
= -\left( 1 - \frac{2Mr}{\Sigma} \right) \dev t^2 - \frac{4M\Kerr r\sin^2\theta}{\Sigma} \dev t \dev \phi
+ \frac{\Sigma}{\Delta}\dev r^2 
\\&~~~~~
+ \Sigma\dev\theta^2 
+ \sin^2\theta \left( r^2+\Kerr ^2+\frac{2M\Kerr^2r\sin^2\theta}{\Sigma} \right) \dev\phi^2,
\end{split}
\label{eq:Kerrmetric}
\end{equation}
where $\Sigma=r^2+\Kerr^2\cos^2\theta$ and $\Delta =r^2-2Mr+\Kerr^2$.
%Note that we adopt the $-$$+$$+$$+$ signature. 
The Weyl scalars are defined as 
\begin{subequations}
\begin{align}
\Psi_0 &= + C_{abcd}l^a m^b l^c m^d,\\
\Psi_1 &= + C_{abcd}l^a m^b l^c n^d,\\
\Psi_2 &= + C_{abcd}l^a m^b \overline{m}^c n^d,\\
\Psi_3 &= + C_{abcd}l^a n^b \overline{m}^c n^d,\\
\Psi_4 &= + C_{abcd}\overline{m}^a n^b \overline{m}^c n^d.
\end{align}
\end{subequations}
%
%$\Psi_0 = +C_{abcd}l^a m^b l^c m^d$ etc.,
where $l^a, {n}^b, {m}^d$ are the Kinnersley tetrad \cite{sfk12}, 
{which are defined in Appendix \ref{sec:def}.}
%The over bar $\overline{m}$ denotes the complex conjugate of $m$. 
In the case of {the} Kerr metric, nonzero Weyl scalar is 
$\Psi_2 = M\rho^3$, where $\rho = -(r-\im \Kerr\cos\theta)^{-1}$.
%$\Psi_2= -{M}/{(r-\im \Kerr\cos\theta)^3}$
The perturbed Weyl scalars are denoted by $\psi_0,~\psi_1,\ldots,~\psi_4$.

We consider the perturbation of the Kerr metric induced by a rotating ring 
which is composed of a set of point particles in circular, geodesic orbit on the equatorial plane. 
%The center of the ring is at the center of the black hole. 
%The ring is given as a set of point masses. Each point mass is on a circular geodesic. 
The energy-momentum tensor of the rotating circular ring, $T^{ab}$, is given as
\begin{equation}
\begin{split}
T^{ab} %&=  \sekibun{}{}{\phi'}\frac{\mass u^a u^b}{u^tr_0{}^2}\delta(r-r_0)\delta(\cos\theta)\delta(\phi-\phi')\\
&= \frac{\mass u^a u^b}{u^tr_0{}^2}\delta(r-r_0)\delta(\cos\theta),
\label{Tring}
\end{split}
\end{equation}
where $r_0$ is the radius of the ring, and 
$u^a = u^t\left( (\partial_t)^a+\Omega(\partial_\phi)^a \right)$
is the four-velocity of the ring.
The angular velocity $\Omega$ and $u^t$ are given as \cite{sfk12}
\begin{equation}
\begin{split}
\Omega &= \frac{\pm M^{1/2}}{r_0{}^{3/2} \pm \Kerr M^{1/2}},
\\
u^t &= \frac{r_0{}^{3/2} \pm M^{1/2} \Kerr}{\sqrt{r_0{}^3 - 3Mr_0{}^2 \pm 2\Kerr M^{1/2}r_0{}^{3/2}}}
,
\end{split}
\end{equation}
{where the upper sign is for prograde rotation and the lower sign is for retrograde rotation.}
The rest mass of the ring becomes $2\pi \mass$.
We assume $\mass \ll M$.

%%
%Without $\partial_t$ and $\partial_\phi$, 
Since our perturbed space-time is independent of $t$ and $\phi$, 
%it is sufficient to consider the case of $\omega=0$ and the $m=0$ mode of 
%the spin-weighted spheroidal harmonics ${}_sS_{lm}^{a\omega}(\theta)$. 
%We 
we expand $\psi_{(s=2)}=\psi_0$ and $\psi_{(s=-2)}=\rho^{-4}\psi_4$ as 
\begin{equation}
\begin{split}
\psi_{(s)}(r,\theta) = \sum_{l=2}^{\infty}R^{(s)}_l(r)~{}_{s}Y_{l}(\theta)
.
%&\psi_0(r,\theta) = \sum_{l=2}^{\infty}R^{(2)}_l(r)~{}_{2}Y_{l}(\theta),
%\\
%& \rho^{-4}\psi_4(r,\theta) = \sum_{l=2}^{\infty}R^{(-2)}_l(r)~{}_{-2}Y_{l}(\theta).
\end{split}
\end{equation}
Here we have defined $_sY_l(\theta)$ as $_sY_l(\theta)\equiv{}_sY_{l0}(\theta,0)$.
The Teukolsky equation for $R^{(s)}$ becomes 
\begin{equation}
\begin{split}
\left[ 
\bibun{}{r}{}  \left(\Delta^{s+1}\bibun{}{r}{}  \right)-\Delta^s {(l-2)(l+3)} 
\right] 
R^{(s)}_l = -8\pi T^{(s)}_l \Delta^s
.
\label{dokei0}
\end{split}
\end{equation}
{The source term $T^{(s)}_l$ is given as}
\begin{equation}
\begin{split}
\frac{T^{(2)}_l}{2\pi} &= 
- \hat{T}_{11} \delta(r-r_0) \frac{\sqrt{(l+2)(l-1)(l+1)l}}{2} {}_{0}Y_l(\pi/2) 
\\
&- 2 \hat{T}_{11} \frac{\im \Kerr}{\sqrt{2} r_0} \delta(r-r_0) \frac{\sqrt{(l+2)(l-1)}}{\sqrt{2}} {}_{-1}Y_l(\pi/2) 
\\
&+ 2 \hat{T}_{13} \frac{r_0{}^3}{r^2} \bibun{}{r}{} \delta(r-r_0) \frac{\sqrt{(l+2)(l-1)}}{\sqrt{2}} {}_{-1}Y_l(\pi/2) 
\\
&- \hat{T}_{33} \frac{r_0}{r^3} \bibun{}{r}{}\left( r^4 \bibun{}{r}{} \delta(r-r_0) \right) {}_{-2}Y_l(\pi/2)
,
\end{split}
\end{equation}
\begin{equation}
T^{(-2)}_l = \frac{\Delta^2}{4} T^{(2)}_l 
,
\end{equation}
where $\hat{T}_{\mu\nu}$ are constants defined as
$\hat{T}_{\mu\nu}=m u_\mu u_\nu/(u^t r_0{}^2)$.

%A simple relation $\frac{\Delta^2}{4}T^{(2)}_l(r)=T^{(-2)}_l(r)$ is hold because of the symmetries.
%WeylスカラーがTeukolskyのそれらと一致しているから、Teukolskyが書いた通りのTeukolsky equationを変えずに使う。
%ソースはエネルギー運動量テンソルとテトラドで作られ、conventionには依存しない。
%Teukolsky72と73でsource termのsignが異なっているのは謎だが、\cite{kfw}にならって73のものを採用する。

The Teukolsky equations are solved by using the Green's function which is given as 
\begin{equation}
%\begin{split}
G_l^{(s)}(r,r') = \frac{ (\Delta \Delta')^{-{s}/{2}} ~ P_l^2(x'_<) Q_l^2(x'_>)}{\sqrt{M^2-\Kerr^2} (l+2)(l+1)l(l-1)},
%\\
%&G_l^{(2)}(r,r') = \frac{P_l^2(x'_<) Q_l^2(x'_>)}{\sqrt{M^2-a^2} \Delta \Delta' (l+2)(l+1)l(l-1)},\\
%&G_l^{(-2)}(r,r') = \frac{\Delta \Delta' P_l^2(x'_<) Q_l^2(x'_>)}{\sqrt{M^2-a^2} (l+2)(l+1)l(l-1)},
%\end{split}
\end{equation}
where $\Delta'=r'^2-2Mr'+a^2$ and
\begin{equation}
x'_{<}\equiv \frac{{\rm min}(r,r')-M}{\sqrt{M^2-\Kerr^2}}
,~~~
x'_{>}\equiv \frac{{\rm max}(r,r')-M}{\sqrt{M^2-\Kerr^2}}.
\end{equation}
A simple relation $(\Delta \Delta')^2 G_l^{(2)}(r,r') = G_l^{(-2)}(r,r')$ holds because of symmetries.
As a result, we obtain
\begin{equation} 
\begin{aligned}
& R_l^{(2)} = \\
& -8\pi^2 \hat{T}_{11} \Delta_0{}^2 G_l^{(2)}(r,r_0) \sqrt{(l+2)(l-1)(l+1)l} ~{_0Y_l(\pi/2)}
\\
& 
-16\sqrt{2}\pi^2 \hat{T}_{11} \frac{\im \Kerr \Delta_0{}^2}{\sqrt{2}r_0} G_l^{(2)}(r, r_0) 
\sqrt{(l+2)(l-1)} ~{_{-1}Y_l(\pi/2)}
\\
& 
-16\sqrt{2}\pi^2 \hat{T}_{13} r_0{}^3 \bibun{}{r_0}{}\left( G_l^{(2)}(r,r_0) \frac{\Delta_0{}^2}{r_0{}^2} \right) 
\\
& \quad \times \sqrt{(l+2)(l-1)} ~{_{-1}Y_l(\pi/2)}
\\
& 
-16\pi^2 \hat{T}_{33} r_0 
\bibun{}{r_0}{}\left[ r_0{}^4 \bibun{}{r_0}{} \left( G_l^{(2)}(r,r_0)\frac{\Delta_0{}^2}{r_0{}^2} \right) \right] ~{_{-2}Y_l(\pi/2)},
\end{aligned}
\end{equation}
\begin{equation}
R_l^{(-2)} = \frac{\Delta^2}{4} R_l^{(2)} ,
\end{equation}
where $\Delta_0=r_0{}^2-2Mr_0+a^2$.

As in the Schwarzschild case, 
we have the symmetry about the equatorial plane,
\begin{equation}
\begin{split}
& \Re(\psi_{0/4}(r, \pi-\theta)) = \Re(\psi_{0/4}(r, \theta)),\\
& \Im(\psi_{0/4}(r, \pi-\theta)) = -\Im(\psi_{0/4}(r, \theta)).
\label{oddeven}
\end{split}
\end{equation}

\begin{figure*}[htb]
\begin{center}
\includegraphics[width=8cm]{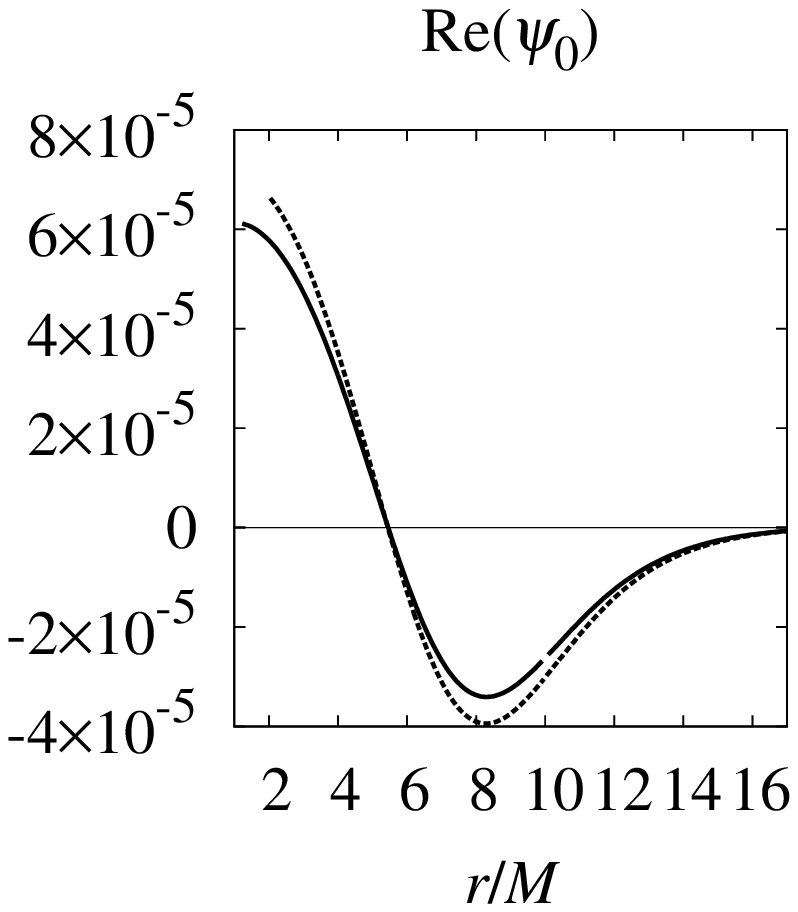}
\includegraphics[width=8cm]{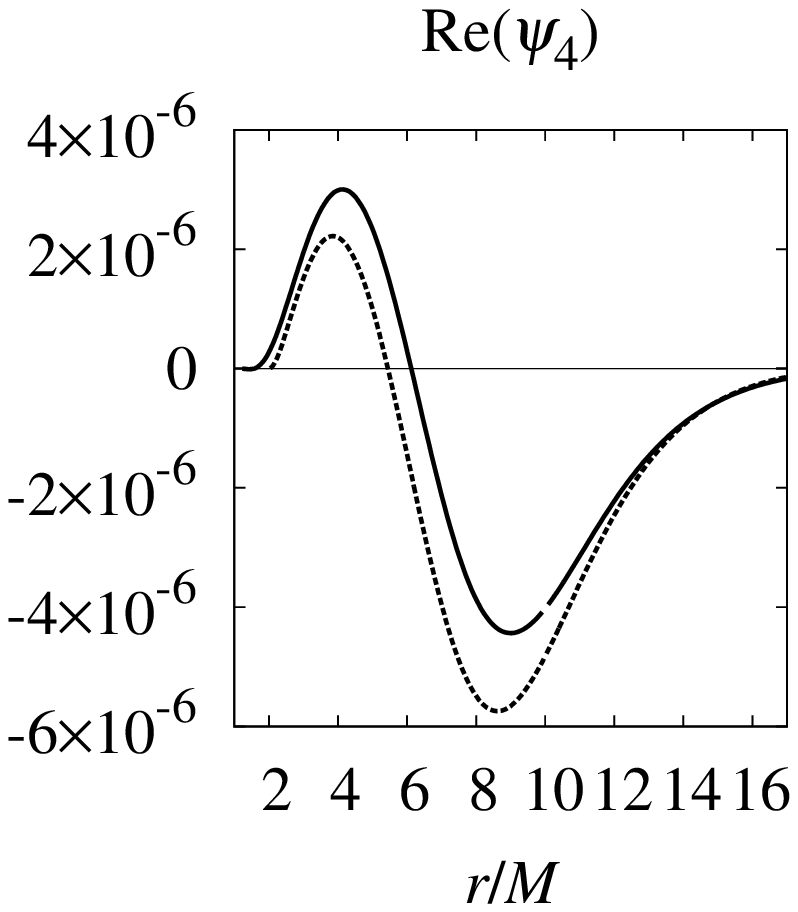}
\\
\includegraphics[width=8cm]{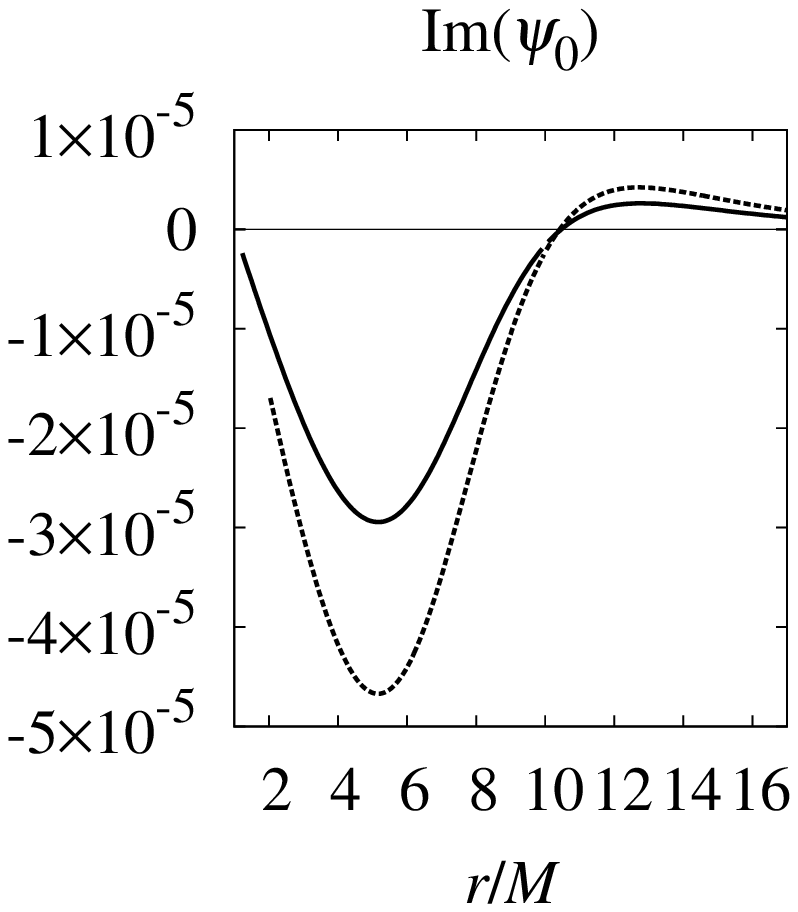}
\includegraphics[width=8cm]{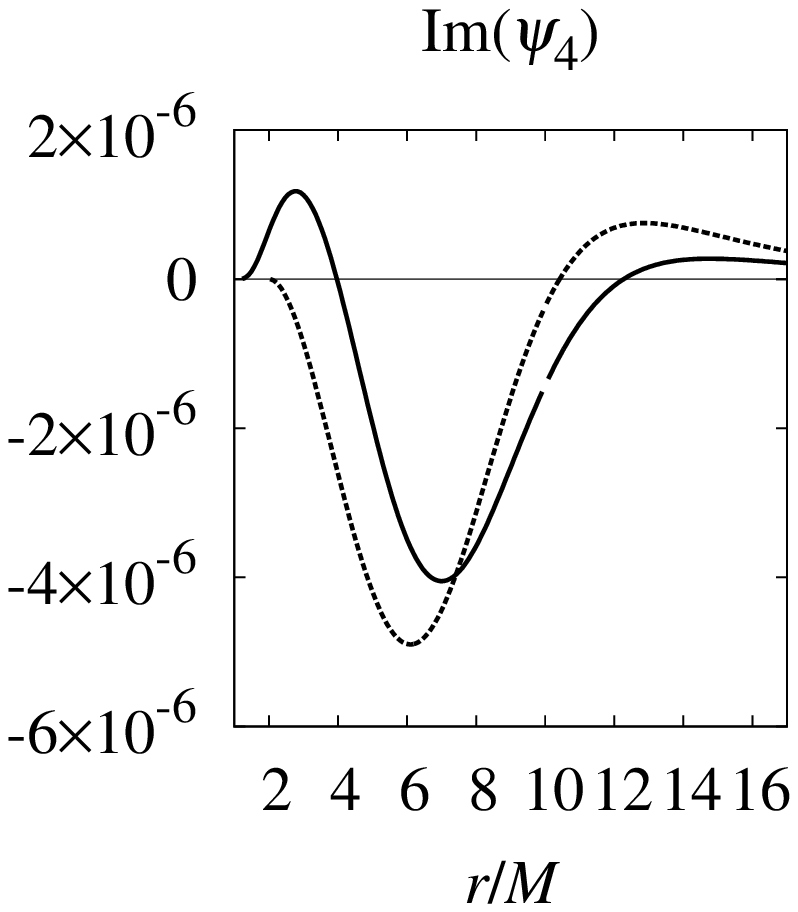}
\end{center}
\caption{
{
Radial dependence of the Weyl scalars $\psi_0$ and $\psi_4$, 
at $\theta=\pi/4$ which are obtained by solving the Teukolsky equations.
The radius of the ring is $r_0=10M$.
Solid lines are for the Kerr case with $\Kerr=0.99M$. 
{The ring's rotation is prograde.}
Dashed lines are for the Schwarzschild case, $\Kerr=0$. 
%We can see the continuity of the fields at $r=r_0$. 
}}
\label{fig:psi04}
\end{figure*}

{The plots of $\psi_0$ and $\psi_4$ are shown in Fig. \ref{fig:psi04}, 
for the case }
{$r_0=10M$, $a=0.99M$, and the ring's rotation is prograde.}

\section{Construction of the perturbed gravitational fields}
\label{sec:construction}

\subsection{The Hertz potential}
\label{sec:hertz}
Next, we use these solution $\psi_0$ and $\psi_4$ in the CCK formalism to find the Hertz potential.
The {ingoing radiation gauge (IRG)}  is defined as $h_{ab}l^b=h^a{}_a=0$. 
The perturbed metric $h_{ab}$ in IRG is related to the Hertz potential as \cite{c}
\begin{equation}
\begin{split}
h_{ab} 
&= -\big[ 
l_a l_b 
(\overline{\pmb\delta} +\alpha +3\overline{\beta} -\overline{\tau})
(\overline{\pmb\delta} +4\overline{\beta} +3\overline{\tau})\overline{\Psi}\\
&~~~~~ ~~~ 
-l_{(a}\overline{m}_{b)}
(\pmb D +\rho -\rhocc)(\overline{\pmb\delta} +4\overline{\beta} +3\overline{\tau})\overline{\Psi} \\
&~~~~~ ~~~
-l_{(a}\overline{m}_{b)}
(\overline{\pmb\delta} -\alpha +3\overline{\beta} -\pai -\overline{\tau})
(\pmb D +3\rhocc)\overline{\Psi} \\
&~~~~~ ~~~ ~~~ 
+\overline{m}_a\overline{m}_b
(\pmb D -\rhocc)(\pmb D +3\rhocc)\overline{\Psi} \big] \\
&~~~ + \left[ {\rm c.c.} \right],
\label{irg}
\end{split}
\end{equation}
%
%この公式はCCK(CK)のものと同じ。
where $\left[ {\rm c.c.} \right]$ represents the complex conjugate of the first term. 
The bold greek characters are derivative operators defined as
\begin{equation}
\begin{split}
& {\pmb D} = l^a\partial_a ,
~~~~~
{\pmb \Delta} = n^a\partial_a ,\\
& {\pmb \delta} = m^a\partial_a ,
~~~~~
\overline{\pmb \delta} = \overline{m}^a\partial_a ,
\end{split}
\end{equation}
and the overline denotes the complex conjugate. 
$\rho$, $\mu$, $\alpha$, $\beta$, $\gamma$, $\pai$, and $\tau$
here are the spin coefficients, {which are defined in Appendix \ref{sec:def}.} 
The Hertz potential $\Psi$ in IRG satisfies the source-free Teukolsky equation with $s=-2$. 
\begin{equation}
( {\pmb \Delta} - 2\mu + 2\gamma ) {\pmb D} \Psi 
+ 3\rho \partial_t \Psi
= ( \overline{\pmb \delta} - 2\overline{\beta} - \overline{\tau} ) ( {\pmb \delta} + 4\beta) \Psi.
\label{IRGhomo2}
\end{equation}

In IRG, since the perturbed space-time is stationary and axisymmetric,
we have
\begin{eqnarray}
&&\psi_0 =
\frac{1}{2}\left(\frac{\partial}{\partial r}\right)^4\overline{\Psi},
\label{psi0-hertz}
\\
&& 4\rho^{-4}\psi_4  =
\frac{1}{2}\sin^2\theta \left(\frac{\partial}{\partial \cos\theta}\right)^4 \sin^2\theta \overline{\Psi} .
\label{psi4-hertz}
\end{eqnarray}
Our task is to find Hertz potential which satisfies \eref{psi0-hertz}, \eref{psi4-hertz} and \eref{IRGhomo2}. 
By substituting the solution of the Teukolsky equation into the left hand side of \eref{psi4-hertz}, 
$\overline{\Psi}$ can be integrated as
\begin{equation}
\begin{split}
\overline{\Psi}(r,\theta) &= \overline{\Psi}_{\rm P}+\overline{\Psi}_{\rm H}, 
\label{hertzPH}
\end{split}
\end{equation}
where
\begin{equation}
\overline{\Psi}_{\rm P} 
= \sum_{l=2}^{\infty} \overline{R_l^{\rm P}}(r) _{2}Y_l(\theta)
\equiv \sum_{l=2}^{\infty} \frac{8{R_l^{(-2)}(r) _{2}Y_l(\theta)}}{(l+2)(l-1)(l+1)l} ,
\label{PsiP}
\end{equation}
\begin{equation}
\begin{split}
\overline{\Psi}_{\rm H} &\equiv \frac{2A}{\sin^2\theta} \bigg( \frac{a(r)}{6}\cos^3\theta +\frac{b(r)}{2}\cos^2\theta
\\&~~~~~~~~~~~~~~~~~~~~~~~~~
+ c(r)\cos\theta +d(r) \bigg)
.
\end{split}
\label{PsiHabcd}
\end{equation}
$A$ is a constant defined as $A\equiv {\mass}/({r_0\sqrt{\Delta_0}})$.
It can be shown that $\Psi_{\rm H}$ is a homogeneous solution of \eref{psi0-hertz} and \eref{psi4-hertz}, 
and satisfies \eref{IRGhomo2} when 
\begin{equation}
\begin{split}
& a(r) = a_1 r^2 (r-3M) + 3a_1\Kerr^2 r + a_2,\\
& b(r) = b_1 (r^2-\Kerr^2) + b_2 (r-M),\\
& c(r) = -\frac{a_1}{2}(r^2+4M^2)(r-M) - \frac{a_2}{2} - \left( c_1 + \frac{a_1}{2}M \right) \Kerr^2
\\&~~~~~~~~~~~~~~~ 
+c_1 r^2+c_2(r-M),\\
& d(r) = \frac{b_1}{2}r^2 + \frac{b_2}{2}r + d_1 (r^3 - 3Mr^2 + 3\Kerr^2 r) + d_2.
\label{abcd}
\end{split}
\end{equation}
Here $a_1$, $a_2$, etc. are arbitrary complex constants, 
and $\Kerr$ is the Kerr parameter.

We see that $\Psi_{\rm P}$ 
satisfies \eref{psi0-hertz}, \eref{psi4-hertz}, and \eref{IRGhomo2} 
everywhere except for $r=r_0$. 
At $r=r_0$, $\Psi_{\rm P}$ has the singularity. 
However, this singularity can be removed by appropriately choosing the parameters in $\Psi_{\rm H}$. 
{
Here, we explain how to obtain those parameters. 
}

First we demand that the metric perturbation and the Weyl scalars should not diverge 
at $\theta=0$ and $\theta=\pi$. These conditions are satisfied if the Hertz potential 
$\Psi$ does not diverge at $\theta=0$ and $\pi$. 
We obtain
\begin{equation}
\begin{split}
& 3 d_1 = \pm a_1,~~~\pm c_1= \pm Ma_1-b_1,
\\
& \pm c_2= \pm\left( 2M^2 - \frac{3\Kerr^2}{2} \right) a_1-b_2,
\\
& 6d_2 = \pm 2a_2 - 3\Kerr^2 b_1 - 3Mb_2.
\label{frompoles}
\end{split}
\end{equation}
where the upper sign is for $\theta=0$, and the lower sign is for $\theta=\pi$. 
We see that these equations are satisfied if and only if 
$a_1=a_2=b_1=b_2=c_1=c_2=d_1=d_2=0$,  i.e.,  $\Psi_{\rm H}=0$.
This implies that we can not obtain the regular solution globally. 
Following \cite{SanoTagoshi14}, we divide the space-time into three regions: 
$I$~:~$(M+\sqrt{M^2-\Kerr^2}<r<r_0)$, $N$~:~$(r>r_0, ~0\leq \theta<\pi/2)$, and $S$~:~$(r>r_0, ~\pi/2<\theta\leq \pi)$.
We set $\Psi_{\rm H}=0$ in the inner region $I$, 
and look for the set of parameters which satisfy \eref{frompoles} in $N$ and $S$, respectively. 
Since there are four equations among eight unknown parameters, the {number of} parameters we have to determine {is} four. 
We adopt $a_1, a_2, b_1, b_2$ in region $N$ as independent variables. 
The parameters in $S$ can be determined by using the symmetry about the equatorial plane \eref{oddeven}. 
We determine the value of these parameters numerically 
by using four continuity conditions for $\psi_1$, $\psi_2$, $h_{33}$, and $\Psi$ 
at $r=r_0$. The necessary equations for $\psi_1$, $\psi_2$ and $h_{33}$ are given in Appendix 
\ref{sec:weyl}. 
{The parameters obtained are shown in Tables \ref{table:a} and \ref{table:b}}
for the case, $M=1$, $m=M/100$, $r_0=10M$, 
and for various value of the Kerr parameter.

\begin{table}[htb]
\caption{The parameters $a_1$ and $a_2$ in $N$. 
This is for the case, $M=1$, $m=M/100$, $r_0=10M$. }
\label{table:a}
\begin{tabular}{ccccc}
\hline \hline
$a/M$ & $\Re(a_1)$ & $\Im(a_1)$ & $\Re(a_2)$ & $\Im(a_2)$ \\ \hline
0.3   & 0.299239e-3 & -3.76054 & -106.766 & -1875.23  \\
0.6   & 0.351477e-3 & -3.29746 & -81.5775 & -1630.99  \\ 
0.9   & 0.317026e-3 & -2.85623 & -59.3722 & -1393.31  \\ 
0.99 & 0.148234e-3 & -2.72761 & -53.4299 & -1323.53  \\ 
\hline \hline
\end{tabular}
\end{table}
\begin{table}[htb]
\caption{The parameters $b_1$ and $b_2$ in $N$. 
This is for the case, $M=1$, $m=M/100$, $r_0=10M$.}
\label{table:b}
\begin{tabular}{ccccc}
\hline \hline
$a/M$ & $\Re(b_1)$ & $\Im(b_1)$ & $\Re(b_2)$ & $\Im(b_2)$ \\ \hline
0.3   & 64.5010 & 30.0819 & -710.120 & 0.706601 \\
0.6   & 62.1322 & 26.3831 & -683.362 & 2.34294 \\ 
0.9   & 60.0441 & 22.8533 & -658.481 & 4.58293 \\ 
0.99 & 59.4624 & 21.8145 & -651.262 & 5.40521 \\ 
\hline \hline
\end{tabular}
\end{table}

\subsection{Perturbed Weyl scalars and the metric perturbation}
\label{sec:results}
\begin{figure*}[htb]
\begin{center}
\includegraphics[width=5.6cm]{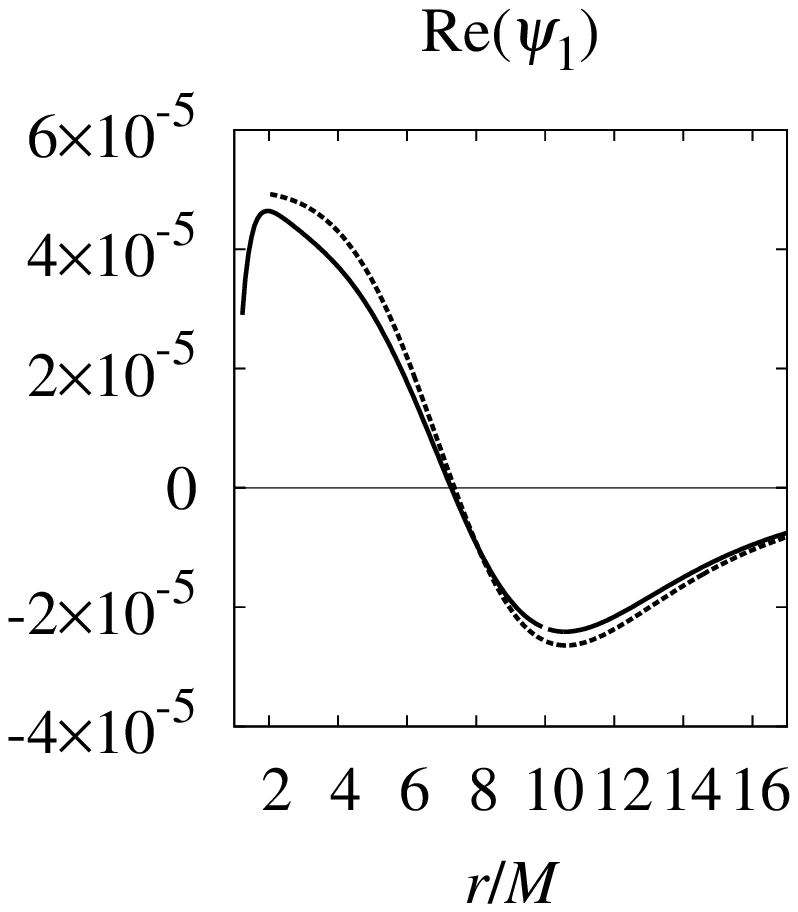}
\includegraphics[width=5.6cm]{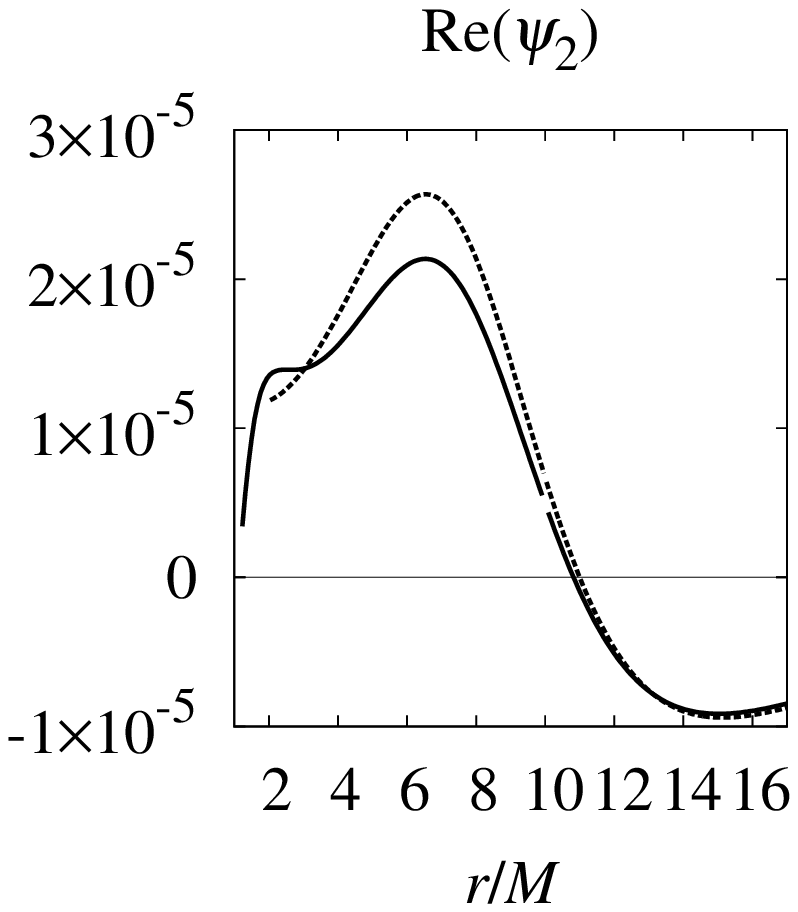}
\includegraphics[width=5.6cm]{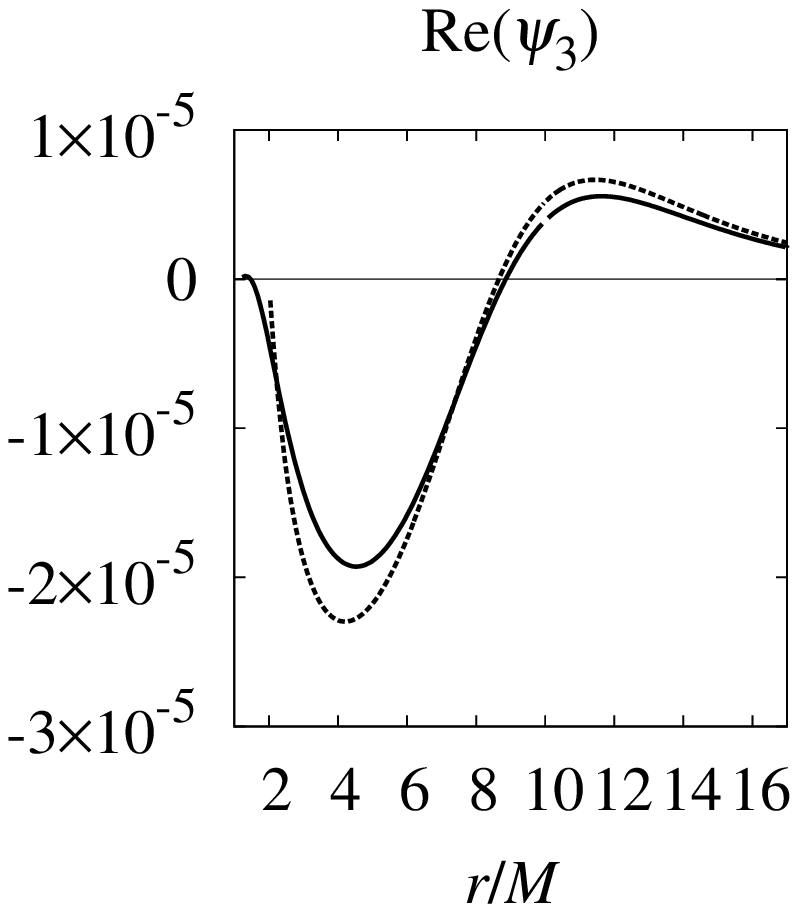}
\\
\includegraphics[width=5.6cm]{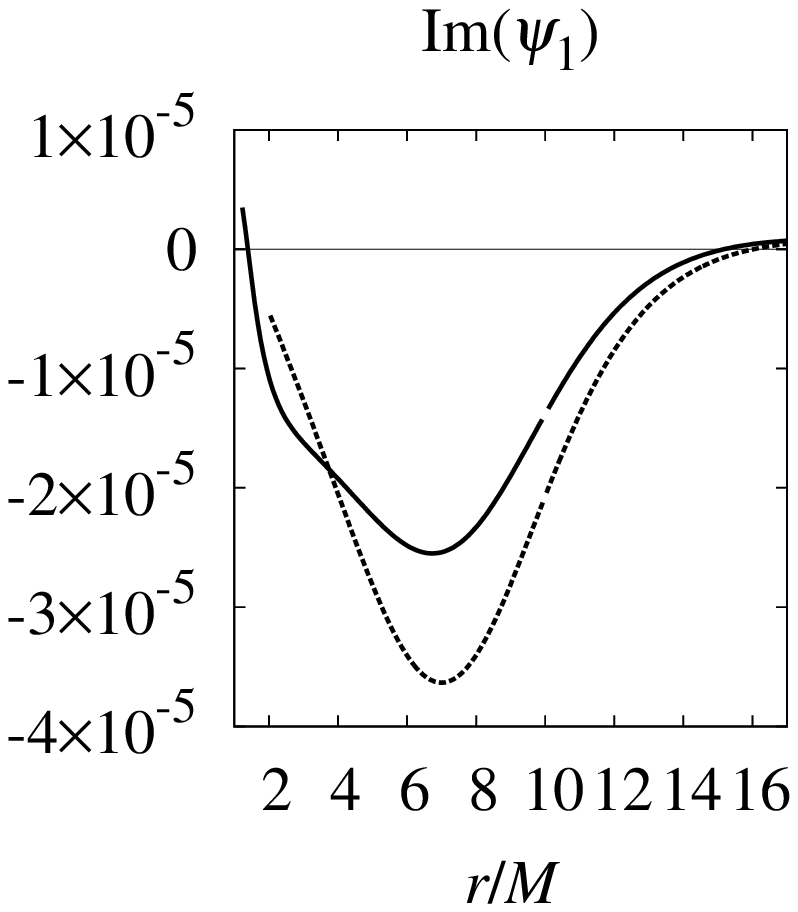}
\includegraphics[width=5.6cm]{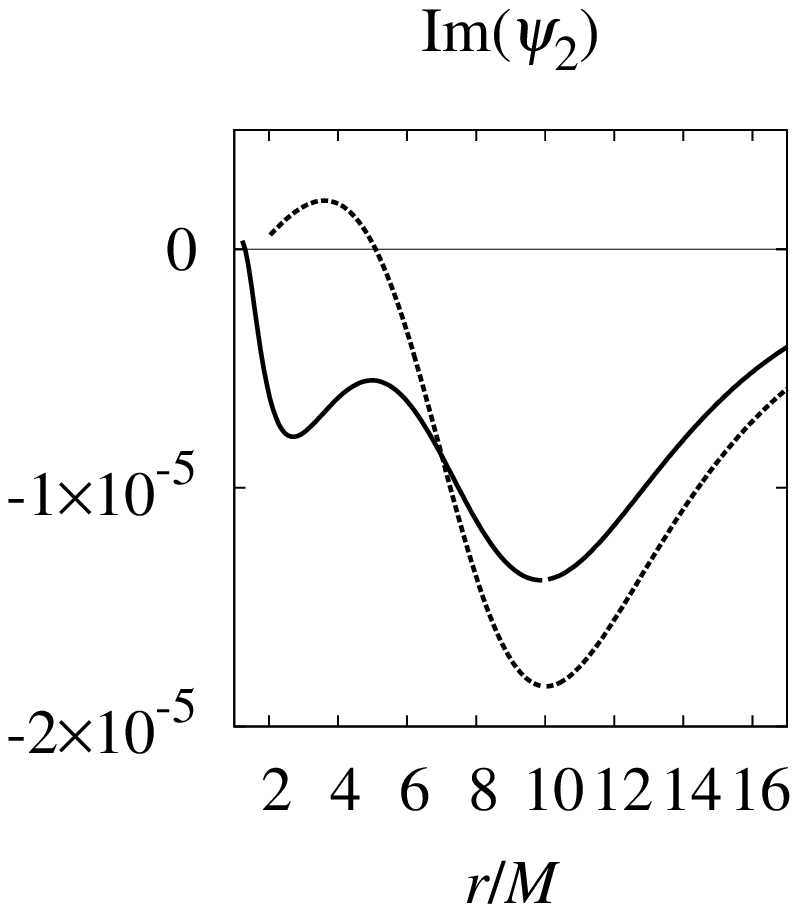}
\includegraphics[width=5.6cm]{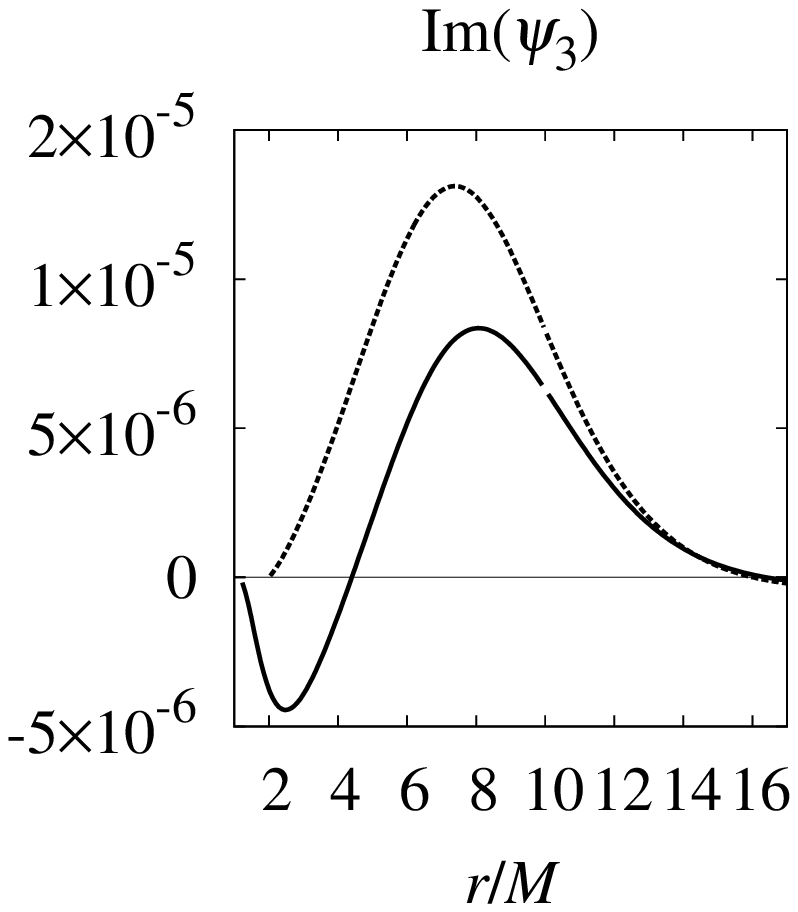}
\end{center}
\caption{
{
Radial dependence of the Weyl scalars $\psi_1$, $\psi_2$, and $\psi_3$, 
at $\theta=\pi/4$ which are derived by using $\Psi=\Psi_{\rm P}+\Psi_{\rm H}$. 
The radius of the ring is $r_0=10M$.
Solid lines are for the Kerr case with $\Kerr=0.99M$. 
{The ring's rotation is prograde.}
Dashed lines are for the Schwarzschild case, $\Kerr=0$. 
We can see the continuity of the fields at $r=r_0$. 
%The difference between the Schwarzschild case and the Kerr case is 
%large near the horizen inside the ring radius. 
}}
\label{fig:psi123}
\end{figure*}

\begin{figure}[htb]
\begin{center}
\includegraphics[width=5.6cm]{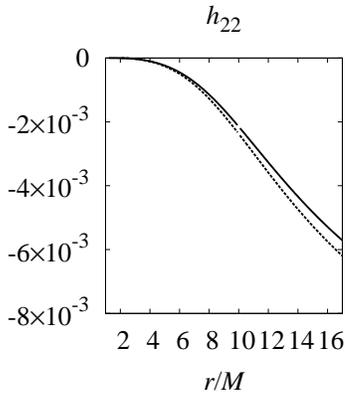}
\end{center}
\caption{
{Radial dependence of $h_{22}$ at $\theta=\pi/4$
which are derived by using $\Psi=\Psi_{\rm P}+\Psi_{\rm H}$. 
The radius of the ring is $r_0=10M$ and the Kerr parameter is $\Kerr=0.99M$. 
The radius of the ring is $r_0=10M$.
Solid lines are for the Kerr case with $\Kerr=0.99M$. 
{The ring's rotation is prograde.}
Dashed lines are for the Schwarzschild case, $\Kerr=0$. 
We can see the continuity of the fields at $r=r_0$. 
}}
\label{fig:h22}
\end{figure}

\begin{figure*}[htb]
\begin{center}
\includegraphics[width=5.6cm]{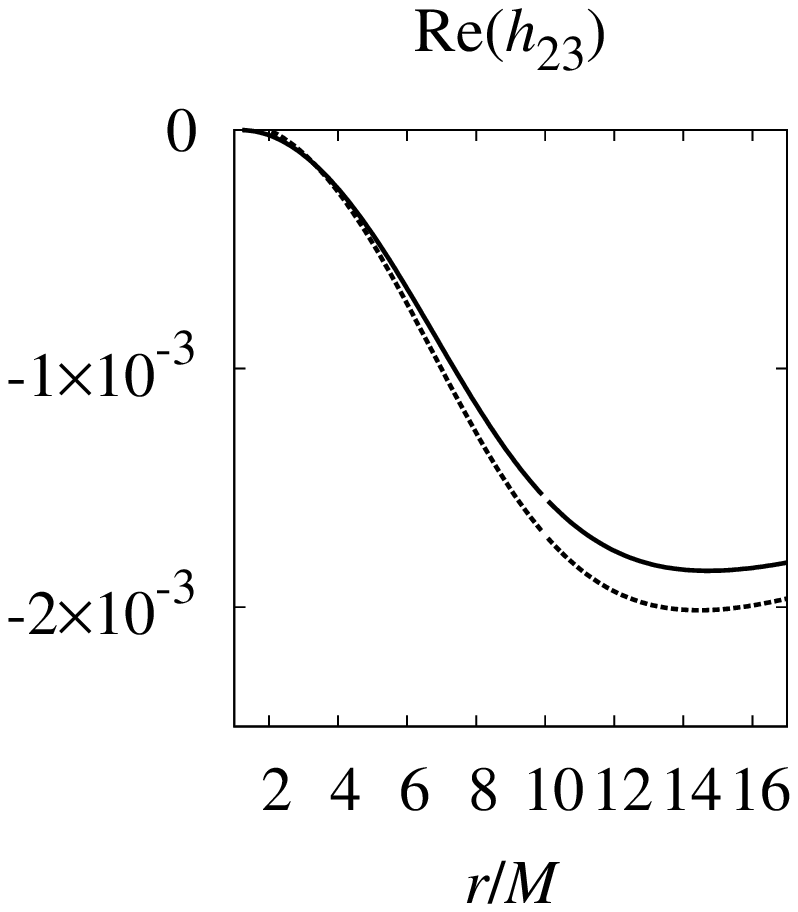}
\includegraphics[width=5.6cm]{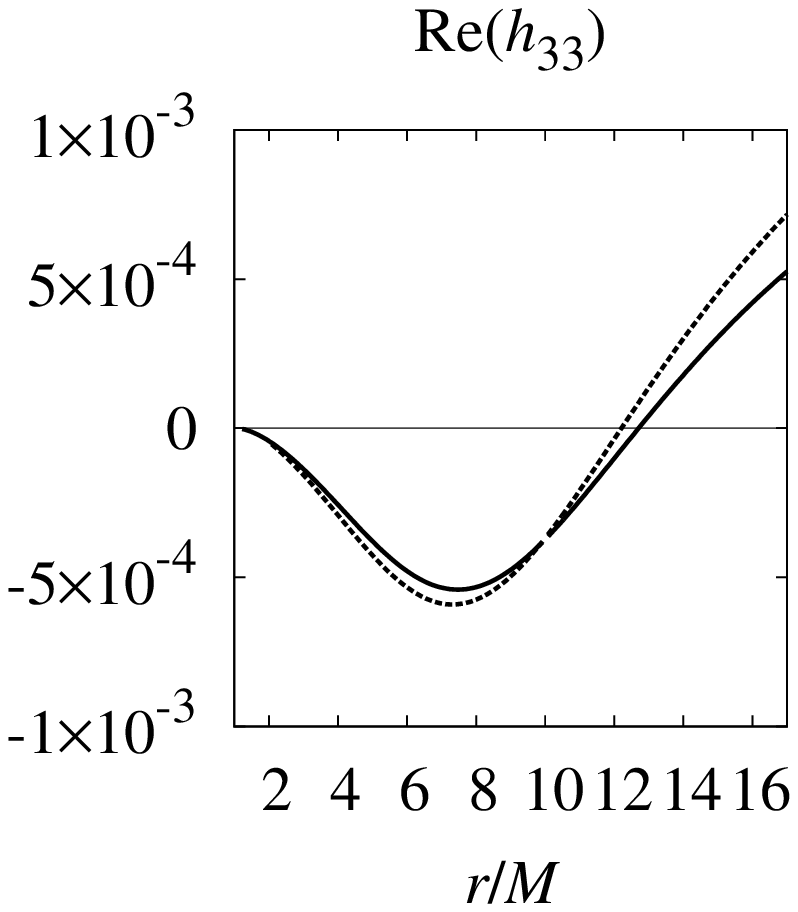}
\includegraphics[width=5.6cm]{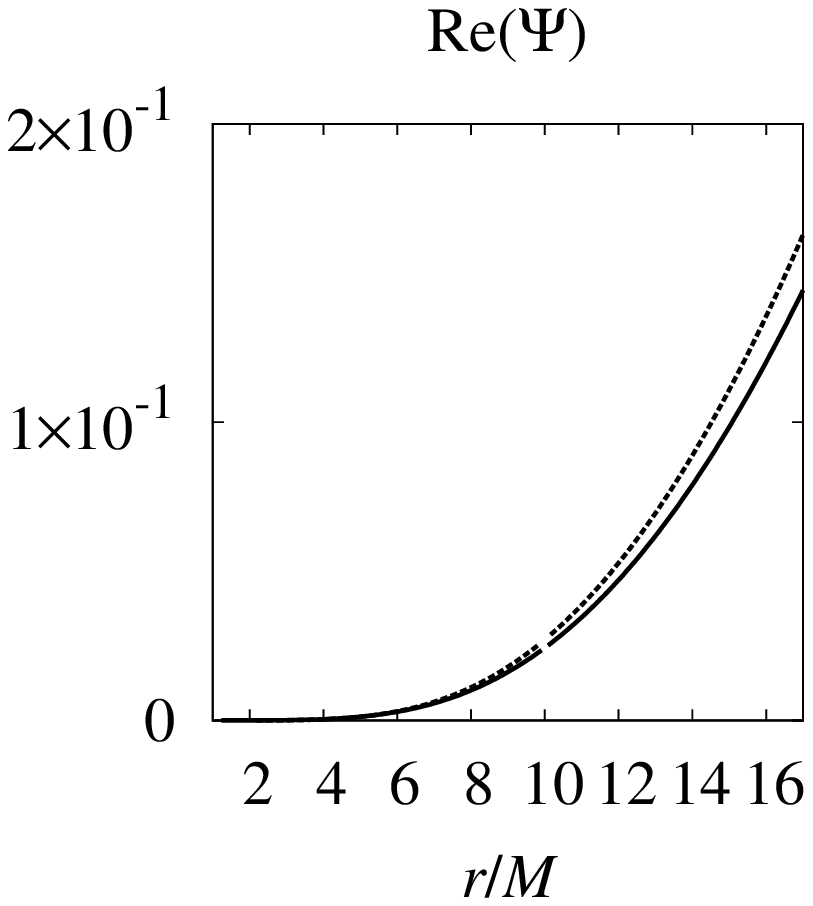}
\\
\includegraphics[width=5.6cm]{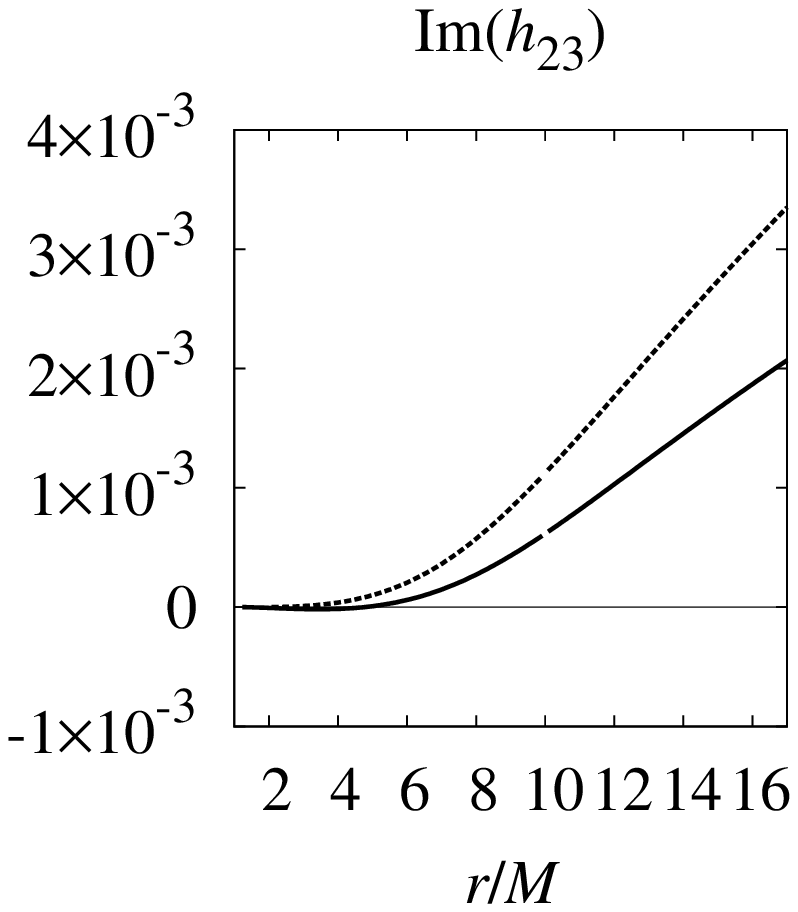}
\includegraphics[width=5.6cm]{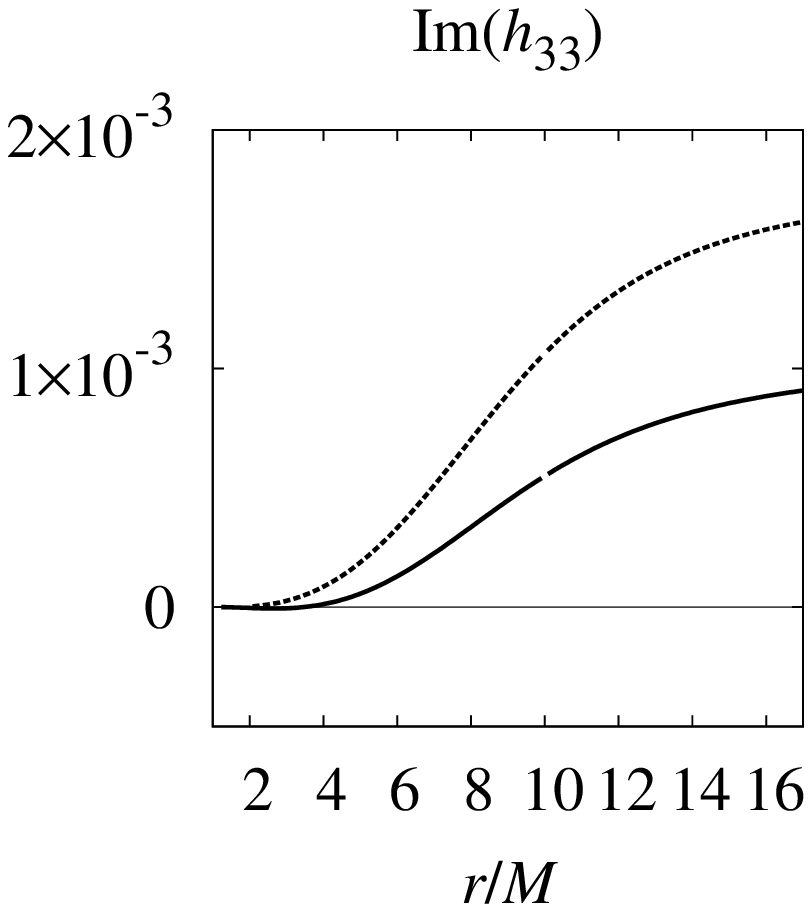}
\includegraphics[width=5.6cm]{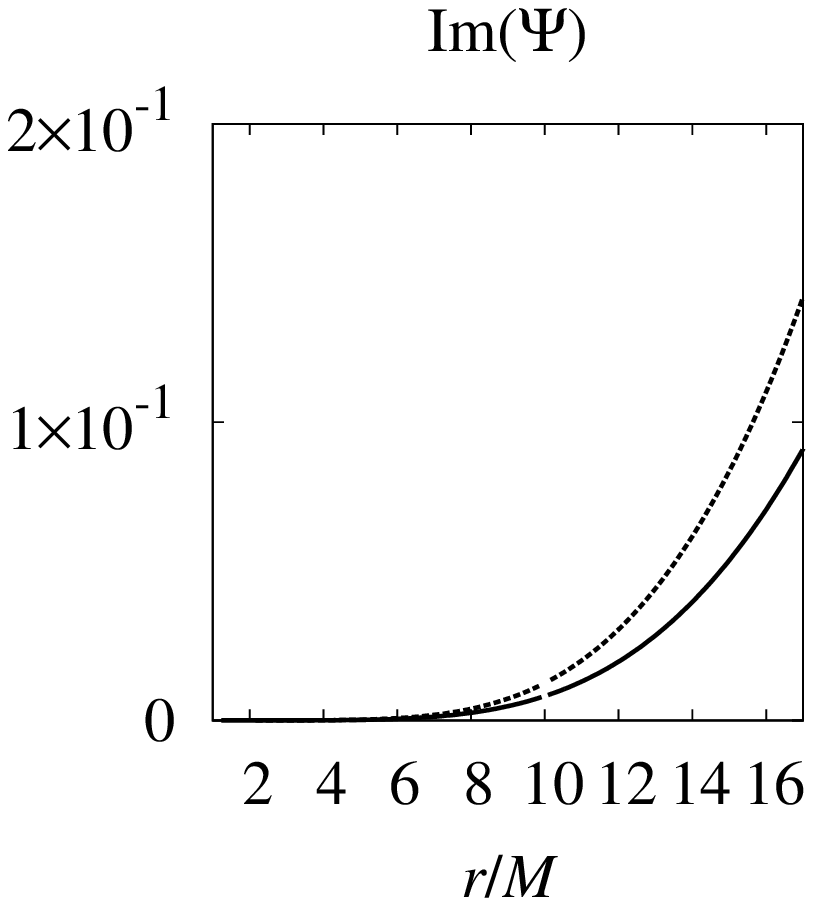}
\end{center}
\caption{
{Radial dependence of the metric perturbations $h_{23}$ and $h_{33}$
derived by using $\Psi=\Psi_{\rm P}+\Psi_{\rm H}$, 
and the Hertz potential $\Psi$ at $\theta=\pi/4$. 
The radius of the ring is $r_0=10M$ and the Kerr parameter is $\Kerr=0.99M$.
{The ring's rotation is prograde.}
Solid lines are for the Kerr case with $\Kerr=0.99M$. 
Dashed lines are for the Schwarzschild case, $\Kerr=0$. 
}}
\label{fig:handHertz}
\end{figure*}

The plots of the Weyl scalars $\psi_1$, $\psi_2$, and $\psi_3$ 
derived by using $\Psi=\Psi_{\rm P}+\Psi_{\rm H}$
are shown in Fig. \ref{fig:psi123}. 
The plots of the metric perturbations $h_{22}$, $h_{23}$, $h_{33}$, and the Hertz potential $\Psi$
are shown in Figs. \ref{fig:h22} and  \ref{fig:handHertz}.
All plots are for the case of the Kerr parameter $\Kerr = 0.99M$.
We find that all of them are continuous at the ring radius, $r=r_0$.

\section{Summary and discussion}
\label{sec:summary}
We derived the perturbed Weyl scalars and the metric perturbation {induced} by a rotating circular 
ring around a Kerr black hole by using the CCK formalism. 
The computation can be done in the same way as in the Schwarzschild case. 
{However, some equations must be derived in the Kerr case.}

In the CCK formalism, the Weyl scalars and the metric perturbation are expressed 
by the Hertz potential $\Psi$ in a radiation gauge. We used the ingoing radiation gauge, and derived 
the $\Psi_{\rm P}$ which has discontinuity on the surface of the sphere at the radius of the ring. 
The homogeneous part of the Hertz potential, $\Psi_{\rm H}$, contains the lower multipole modes $l=0, 1$
and some gauge freedom.
We derived {the general form of} $\Psi_{\rm H}$ in the Kerr case, \eref{PsiHabcd}, \eref{abcd}. 
As in the Schwarzschild case, the lower modes and gauge freedom appear as eight complex
parameters.
We determined {these parameters in} $\Psi_{\rm H}$ numerically by demanding the continuity of 
{the Weyl scalars, metric perturbation, and} the Hertz potential at the ring radius.
{The expressions for $\psi_1$ and $\psi_2$ in terms of the Hertz potential 
\eref{psi1-hertz}, \eref{psi2-hertz} are derived in the Kerr case, and are used to impose the continuity condition.}
We obtained the Hertz potential which is smooth except for the equatorial plane outside the ring radius. 
The perturbed Weyl scalars, $\psi_1$, $\psi_2$, $\psi_3$, and the metric perturbation 
also contain discontinuity on the equatorial plane outside the ring. 
This is completely the same as in the Schwarzschild case in \cite{SanoTagoshi14}.
Note that, as in the Schwarzschild case, 
in the determination of the parameters in $\Psi_{\rm H}$, 
we did not use the relation between the mass and the angular momentum perturbation, 
and the parameters in $\Psi_{\rm H}$. We only needed to use the continuity condition.

{
One of the most important extension of this work is the case of a particle 
orbiting around a Kerr black hole
\cite{LoustoWhiting, ori03, BarackOri2001, ksf+10, skf+11, pmb}. 
Since the problem becomes nonstationary in such a case,
the Teukolsky equation and the spin-weighted spheroidal harmonics must be solved numerically. 
Thus, the problem becomes much more complicated. 
But once we obtain the gravitational field in a radiation gauge,
it will be possible to compute the gravitational self-force acting on the point particle 
by using the prescription by Pound {\it et al.} \cite{pmb}. 
We want to work on this problem in the future. }

\acknowledgments
YS's work was supported in part by 
{Graduate School of Science, Osaka University}.
HT's work was supported in part by Grant-in-Aid for Scientific Research
(C) No. 23540309, Grant-in-Aid for Scientific Research (A) No. 24244028, 
and by MEXT Grant-in-Aid for Scientific Research on Innovative Areas, 
``New Developments in Astrophysics Through Multi-Messenger Observations of Gravitational Wave Sources'', 
Nos. 24103005, 
and by JSPS Core-to-Core Program, A. Advanced Research Networks.

\appendix
\renewcommand\theequation{\Alph{section}.\arabic{equation}}

\section{Definitions of the Kinnersley tetrad and the spin coefficients}\label{sec:def}
The Kinnersley tetrad is defined as
\begin{subequations}\label{kinnersley}
\begin{align}
l^a &= \frac{r^2+\Kerr^2}{\Delta} (\partial_t)^a + (\partial_r)^a + \frac{\Kerr}{\Delta}(\partial_\phi)^a
,\\
n^a &= \frac{\Delta}{2\Sigma} 
\left[ \frac{r^2+\Kerr^2}{\Delta} (\partial_t)^a - (\partial_r)^a + \frac{\Kerr}{\Delta}(\partial_\phi)^a \right]
,\\
m^a &= \frac{1}{\sqrt{2} (r+\im\Kerr \cos\theta)} \nonumber
\\ &\times
\left[ \im\Kerr\sin\theta (\partial_t)^a + (\partial_\theta)^a +\im\csc\theta(\partial_\phi)^a \right]
,\\
\overline{m}^a &= \frac{1}{\sqrt{2} (r-\im\Kerr \cos\theta)} \nonumber
\\ &\times
\left[ -\im\Kerr\sin\theta (\partial_t)^a + (\partial_\theta)^a -\im\csc\theta(\partial_\phi)^a \right]
.
\end{align}
\end{subequations}
{These} vectors are null, and satisfy normalization and orthogonality conditions.
\begin{subequations}
\begin{align}
l_a l^a = n_a n^a = m_a m^a = \overline{m}_a \overline{m}^a =0,
\\
- l_a n^a = m_a \overline{m}^a = 1,
\\
l_a m^a = l_a \overline{m}^a = n_a m^a = n_a \overline{m}^a =0.
\end{align}
\end{subequations}

The Ricci rotation coefficients are defined as
\begin{equation}
\gamma_{\rho\mu\nu} \equiv  (\nabla_b (e_\rho)_a) (e_\mu)^a (e_\nu)^b
= - \gamma_{\mu\rho\nu}
,
\label{RicciRotation}
\end{equation}
where $\nabla_a$ is the covariant derivative, and $(e_1)^a=l^a$, $(e_2)^a=n^a$,
$(e_3)^a=m^a$, and $(e_4)^a=\overline{m}^a$.
With the Kerr metric \eref{eq:Kerrmetric} and the Kinnersley tetrad \eref{kinnersley}, 
non-zero spin coefficients 
{
$\rho$, $\mu$, $\beta$, $\gamma$, $\alpha$, $\pai$, and $\tau$
}
are given as
\begin{subequations}
\begin{align}
&
\gamma^{2}{}_{34} = \rho = -\frac{1}{r-\im \Kerr\cos\theta},
\\&
\gamma^{1}{}_{43} = -\mu = -\frac{ \Delta }{ 2\Sigma } \rho,
\\&
\frac{1}{2}( \gamma^{2}{}_{23} + \gamma^{4}{}_{43} ) = \beta = -\rhocc\frac{\cot\theta}{2\sqrt{2}},
\\&
\frac{1}{2}( \gamma^{2}{}_{22} + \gamma^{4}{}_{42} ) = \gamma = \mu + \frac{ r-M }{ 2\Sigma },
\\&
\frac{1}{2}( \gamma^{2}{}_{24} + \gamma^{4}{}_{44}) = \alpha = \pi - \overline{\beta},
\\&
\gamma^{3}{}_{21} = -\pai = - \rho^2 \frac{ \im \Kerr \sin\theta }{ \sqrt{2} },
\\&
\gamma^{4}{}_{12} = \tau = \frac{\rho}{~\rhocc~}\paicc
.
\label{spin-list}
\end{align}
\end{subequations}
%
%
\if0
\begin{equation}
\begin{split}
&
\rho = -\frac{1}{r-\im\Kerr\cos\theta},~~~
\mu = \frac{\Delta}{2\Sigma} \rho,~~~
\gamma = \mu + \frac{r-M}{2\Sigma},
\\
&
\varpi = \rho^2\frac{\im\Kerr\sin\theta}{\sqrt{2}},~~~
\tau = \frac{~\rho~}{\rhocc}\overline{\varpi},
\\
&
\beta = -\rhocc \frac{\cot\theta}{2\sqrt{2}},~~~
\alpha = \varpi - \overline{\beta}.
\end{split}
\end{equation}
%
\fi

\section{Weyl scalars $\psi_1, \psi_2$ and the metric perturbation $h_{33}$ in terms of the Hertz potential}
\label{sec:weyl}
In the IRG, expressions for the Weyl scalars in terms of the Hertz potential
can be obtained by using \eref{irg} and
\begin{multline}
-2C_{abcd}
= \nabla_d \nabla_b h_{ac} + \nabla_c \nabla_a h_{bd} 
- \nabla_d \nabla_a h_{bc} - \nabla_c \nabla_b h_{ad}
\\
+C^{(0)}_{aecd}h^e{}_b-C^{(0)}_{becd}h^e{}_a,
\label{Ctoh}
\end{multline}
\begin{eqnarray}
h_{\mu\nu;\rho\sigma} 
& \equiv & (\nabla_d \nabla_c h_{ab}) (e_\mu)^a (e_\nu)^b (e_\rho)^c (e_\sigma)^d
\nonumber\\
&=& (h_{\mu\nu,\rho} + 2h_{\kappa(\mu} \gamma^\kappa{}_{\nu)\rho})_{,\sigma} 
\nonumber\\
&& 
+ (h_{\lambda\mu,\rho} 
+2 h_{\kappa(\lambda} \gamma^\kappa{}_{\mu)\rho}) \gamma^\lambda{}_{\nu\sigma}
\nonumber\\
&&
+ (h_{\lambda\nu,\rho}
+2 h_{\kappa(\lambda} \gamma^\kappa{}_{\nu)\rho}) \gamma^\lambda{}_{\mu\sigma}
\nonumber\\
&&
+ (h_{\mu\nu,\lambda}
+2 h_{\kappa(\mu} \gamma^\kappa{}_{\nu)\lambda}) \gamma^\lambda{}_{\rho\sigma} .
\end{eqnarray}
{
where $C_{abcd}$ and $C^{(0)}_{abcd}$ are
the first order perturbation and the unperturbed part of the Weyl tensor, respectively. 
Here the directional derivatives are denoted by $_{,1}$, $_{,2}$, and so on.}
\begin{equation}
\diffl = {}_{,1}, ~~~ \diffn = {}_{,2}, ~~~ \diffm = {}_{,3}, ~~~ \diffmcc = {}_{,4}.
\end{equation}

First, $\psi_1$ is 
\begin{multline}
-2\psi_1 
= 2\rhocc\taucc h_{33} - 2\pai(\diffl + \rho)h_{33} + \diffl \diffl h_{23}
\\
+ \pai\rhocc h_{33} - \diffl (\rhocc h_{23}) - (\diffl - \rho)(\tau h_{33}) .
\label{psi1-h}
\end{multline}
This equation can be reduced to
\begin{multline}
2\psi_1
= \diffl \diffl \diffl (\diffmcc + 4\betacc) \Psicc 
- 3\pai \diffl (\diffl + 2\rho) \diffl \Psicc .
\label{psi1-hertz}
\end{multline}
When $\partial_t\Psi = \partial_\phi\Psi = 0$, we obtain
\begin{multline}
2\psi_1
= \frac{-1}{\sqrt{2}} \henbibun{}{r}{3} \frac{\rho}{\sin^2\theta}\henbibun{}{\theta}{} \sin^2\theta \Psicc 
\\
- 3\pai \henbibun{}{r}{} \frac{1}{\rho^{2}} \henbibun{}{r}{}\rho^2 \henbibun{}{r}{} \Psicc .
\label{psi1-hertz-ring}
\end{multline}
Therefore, by substituting \eref{PsiP} into this,  we obtain
\begin{multline}
\psi_1^{\rm P} = \frac{1}{2} \sum_{l=2}^{\infty}
\left[ -\frac{1}{\sqrt{2}} \henbibun{}{r}{3} \rho \overline{R_l^{\rm P}} \sqrt{(l+2)(l-1)} ~{}_1Y_l
\right.
\\
\left.
- 3\pai \henbibun{}{r}{} \frac{1}{\rho^2} \henbibun{}{r}{} \rho^2 \henbibun{}{r}{} 
\overline{R_l^{\rm P}} ~{}_2Y_l \right] .
\label{psi1P-hertz}
\end{multline}

In a similar manner, $\psi_2$ becomes
\begin{multline}
2\psi_2
= \diffl \diffl \rho (\diffmcc + 2\betacc) \frac{1}{\rho} (\diffmcc + 4\betacc) \Psicc
\\
- 4\pai (\diffl + \rho) \diffl (\diffmcc + 4\betacc) \Psicc
+ 6\pai \diffl \pai \diffl \Psicc .
\label{psi2-hertz}
\end{multline}
When $\partial_t\Psi = \partial_\phi\Psi = 0$, this equation reduces to 
\begin{multline}
2\psi_2
= \frac{1}{2}\henbibun{}{r}{2} \frac{\rho^2}{\sin\theta} 
\henbibun{}{\theta}{} \frac{1}{\sin\theta} \henbibun{}{\theta}{} \sin^2\theta \Psicc
\\
+ \frac{4\pai}{\sqrt{2} \rho} \henbibun{}{r}{} \rho \henbibun{}{r}{} \frac{\rho}{\sin^2\theta}
\henbibun{}{\theta}{} \sin^2\theta \Psicc
\\
+ 6\pai \henbibun{}{r}{} \pai \henbibun{}{r}{} \Psicc .
\label{psi2-hertz-ring}
\end{multline}
By substituting \eref{PsiP} into this,  we obtain
\begin{multline}
\psi_2^{\rm P} = \frac{1}{2} \sum_{l=2}^{\infty}
\left[ \frac{1}{2} \henbibun{}{r}{2} \rho^2 \overline{R_l^{\rm P}} \sqrt{(l+2)(l-1)(l+1)l} ~{}_0Y_l
\right.
\\
+ \frac{4}{\sqrt{2}} \frac{\pai}{\rho} \henbibun{}{r}{} \rho \henbibun{}{r}{} \rho 
 \overline{R_l^{\rm P}} \sqrt{(l+2)(l-1)} ~{}_1Y_l
\\
\left.
+ 6 \pai \henbibun{}{r}{} \pai \henbibun{}{r}{} \overline{R_l^{\rm P}} ~{}_2Y_l
\right] .
\label{psi2P-hertz}
\end{multline}

On the other hand, the metric perturbation $h_{33}^P$ can be derived 
by substituting \eref{PsiP} into \eref{irg}. 
{When $\partial_t\Psi = \partial_\phi\Psi = 0$,} we have 
\begin{equation}
h_{33}^{\rm P} = - \sum_{l=2}^{\infty} \frac{1}{\rhocc^2} \henbibun{}{r}{} \rhocc^2 \henbibun{}{r}{}
\overline{R_l^{\rm P}} ~{}_2Y_l .
\end{equation}

%%%%%%%%%%%%%%%%%%%
%%%%%%%%%%%%%%%%%%%
%%%%%%%%%%%%%%%%%%%
%
%%%%%%%%%%%%%%%%%%%%%%%%%%%%%%%%%%%%%%%


\begin{thebibliography}{99}
\bibitem{eLISA} P. Amaro-Seoane, S. Aoudia, S. Babak, P. Binetruy, 
E. Berti, A. Bohe, C. Caprini, M. Colpi, N. J. Cornish, K. Danzmann, et al., arXiv:1201.3621.
\bibitem{SNK}
N. Seto, S. Kawamura and T. Nakamura, Phys. Rev. Lett. {\bf 87}, 221103 (2001).
\bibitem{DECIGO}
S. Kawamura et al., Class. Quant. Grav., {\bf 28},  094011 (2011).
\bibitem{BBO}
J. Crowder, N. J. Cornish, Phys. Rev. D {\bf 72}, 083005 (2005).
\bibitem{SanoTagoshi14}
Y. Sano and H. Tagoshi,
Phys. Rev. D {\bf 90}, 044043 (2014).
\bibitem{RW1957}
Regge and Wheeler, Phys. Rev. {\bf 108}, 1063 (1957).
\bibitem{Z1970}
Zerilli, Phys. Rev. D {\bf 2}, 2141 (1970).
\bibitem{c}
P. L. Chrzanowski, 
Phys. Rev. D {\bf 11}, 2042 (1975).
\bibitem{ck}
J. M. Cohen and L. S. Kegeles, 
Phys. Rev. D {\bf 10}, 1070 (1974).
\bibitem{ck75}
J. M. Cohen and L. S. Kegeles,
Phys. Lett. {\bf 54A}, 5 (1975).
\bibitem{kc}
L. S. Kegeles and J. M. Cohen, 
Phys. Rev. D {\bf 19}, 1641 (1979).
\bibitem{wald1978}
R.~M.~Wald, Phys. Rev. Lett. {\bf 41}, 203 (1978).
\bibitem{Stewart1979}
J. M. Stewart, 
Royal Society of London Proceedings Series A {\bf 367}, 527 (1979).
\bibitem{LoustoWhiting}
C. O. Lousto and B. F. Whiting, 
Phys. Rev. D {\bf 66}, 024026 (2002).
\bibitem{ori03}
A. Ori, Phys. Rev. D {\bf 67}, 124010 (2003).
%\bibitem{eLISA} P. Amaro-Seoane, S. Aoudia, S. Babak, P. Binetruy, 
%E. Berti, A. Bohe, C. Caprini, M. Colpi, N. J. Cornish, K. Danzmann, et al., arXiv:1201.3621.
%\bibitem{SNK}
%N. Seto, S. Kawamura and T. Nakamura, Phys. Rev. Lett. {\bf 87}, 221103 (2001).
%\bibitem{DECIGO}
%S. Kawamura et al., Class. Quant. Grav., {\bf 28},  094011 (2011).
%\bibitem{BBO}
%J. Crowder, N. J. Cornish, Phys. Rev. D {\bf 72}, 083005 (2005).
\bibitem{YunesGonzalez}
N. Yunes and J. A. Gonz$\acute{\rm a}$lez,
Phys. Rev. D {\bf 73}, 024010 (2006).
\bibitem{kfw}
T. S. Keidl, J. L. Friedman, and A. G. Wiseman, 
Phys. Rev. D {\bf 75}, 124009 (2007).
\bibitem{wald1973}
R.~M.~Wald, J. Math. Phys. {\bf 14}, 1453 (1973).
\bibitem{BarackOri2001}
L. Barack and A. Ori, 
Phys. Rev. D {\bf 64}, 124003 (2001).
\bibitem{ksf+10}
T. S. Keidl, A. G. Shah, J. L. Friedman, D.-H. Kim, and L. R. Price, 
Phys. Rev. D {\bf 82}, 124012 (2010).
\bibitem{skf+11}
A. G. Shah, T. S. Keidl, J. L. Friedman, D.-H. Kim, and L. R. Price, 
Phys. Rev. D {\bf 83}, 064018 (2011).
\bibitem{sfk12}
A. G. Shah, J. L. Friedman, T. S. Keidl, 
Phys. Rev. D {\bf 86}, 084059 (2012).
\bibitem{pmb}
A. Pound, C. Merlin, and L. Barack, 
Phys. Rev. D {\bf 89}, 024009 (2014).
\bibitem{np62}
E. T. Newman and R. Penrose, 
J. Math. Phys. {\bf 3}, 566 (1962). 
\bibitem{teu}
S. A. Teukolsky, 
Phys. Rev. Lett. {\bf 29}, 1114 (1972); Astrophys. J. {\bf 185}, 635 (1973).
\bibitem{will74}
C.~M.~Will, Astrophys. J. {\bf 191}, 521 (1974).
\bibitem{will75}
C.~M.~Will, Astrophys. J. {\bf 196}, 41 (1975).
\bibitem{np66}
E. T. Newman and R. Penrose, 
J. Math. Phys. {\bf 7}, 863 (1966). 
\bibitem{castillo}
G. F. Torres del Castillo, 
REVISTA MEXICANA DE $\acute{\rm F}$ISICA S {\bf 53} (2) 125 (2005).
%
%\bibitem{gravity}
%C. W. Misner, K. S. Thorne, J. A. Wheeler, 
%{\it Gravitation} (Physics Series) (W H Freeman \& Co, 1973).
\bibitem{nok}
T. Nakamura, K. Oohara, and Y. Kojima, 
Prog. Theor. Phys. Suppl. {\bf 90,} (1987).
%\bibitem{waldtext}
%R.~M.~Wald, {\it General Relativity} (The University of Chicago Press, Chicago, 1984).
\end{thebibliography}
\end{document}